# Stochastic Real-Time Economic Dispatch for Integrated Electric and Gas Systems Considering Uncertainty Propagation and Pipeline Leakage

Peiyao Zhao, *Student Member, IEEE*; Zhengshuo Li, *Senior Member, IEEE*; Jiahui Zhang; Xiang Bai; Jia Su

*Abstract*—Gas-fired units (GFUs) with rapid regulation capabilities are considered an effective tool to mitigate fluctuations in the generation of renewable energy sources and have coupled electricity power systems (EPSs) and natural gas systems (NGSs) more tightly. However, this tight coupling leads to uncertainty propagation, a challenge for the real-time dispatch of such integrated electric and gas systems (IEGSs). Moreover, pipeline leakage failures in the NGS may threaten the electricity supply reliability of the EPS through GFUs. To address these problems, this paper first establishes an operational model considering gas pipeline dynamic characteristics under uncertain leakage failures for the NGS and then presents a stochastic IEGS real-time economic dispatch (RTED) model considering both uncertainty propagation and pipeline leakage uncertainty. To quickly solve this complicated large-scale stochastic optimization problem, a novel notion of the coupling boundary dynamic adjustment region considering pipeline leakage failure (LCBDAR) is proposed to characterize the dynamic characteristics of the NGS boundary connecting GFUs. Based on the LCBDAR, a noniterative decentralized solution is proposed to decompose the original stochastic RTED model into two subproblems that are solved separately by the EPS and NGS operators, thus preserving their data privacy. In particular, only one-time data interaction from the NGS to the EPS is required. Case studies on several IEGSs at different scales demonstrate the effectiveness of the proposed method.

## Nomenclature

### A. Sets

| | |
|---|---|
| $\Omega_A / \Omega_W / \Omega_L$ | Sets of automatic generation control (AGC) units, wind farms, and power demands |
| $\Omega_{U,s} / \Omega_{W,s} / \Omega_{L,s}$ | Sets of units, wind farms, and power demands on bus $s$ in the power system |
| $\Omega_S$ | Set of buses in the power system |
| $\Omega_{GW} / \Omega_{CP} / \Omega_{GL} / \Omega_G$ | Sets of gas wells, compressors, gas loads, and gas-fired units in the natural gas system |
| $\mathcal{U}$ | Robust uncertainty model based on sample paths |
| $\mathcal{W}$ | Discrete set including the normal condition and pipeline leakage conditions of the pipeline network |

### B. Parameters

| | |
|---|---|
| $p_{i,t}^w / P_{i,t}^w$ | Practical and forecast generation of wind farm $i$ at time $t$ |
| $c_i^u / c_i^{up} / c_i^{dw} / c_i^r$ | Unit cost of the base output, upward/downward reserve, and regulation generation output of unit $i$ |
| $N$ | Number of sample paths in the sample path set. |
| $D_{i,t}$ | Power demand of load $i$ at time $t$ |
| $R_{sys}^{up} / R_{sys}^{dw}$ | Systemwide upward/downward reserve requirements |
| $L_m^{tr}$ | Transmission capacity of line $m$ |
| $H_{m,s}^{PT}$ | Transmission distribution factor from bus $s$ to line $m$ |
| $P_i^{min} / P_i^{max}$ | Lower/upper bounds of the output of AGC unit $i$ |
| $r_i^{up} / r_i^{dw}$ | Up- and down ramp rates of AGC unit $i$ |
| $l / S / d / \lambda$ | Axial length, cross-sectional area, inner diameter, and friction factor of the gas pipeline |
| $c$ | Sound speed |
| $\pi_z^t / q_z^t / \pi_n^t / q_n^t$ | Pressures and mass flow rates at discretized node $z$ and practical node $n$ in the natural gas system |
| $\mu_k / A_k$ | Leakage coefficient and leakage hole area of the gas pipeline |
| $\kappa / M$ | Adiabatic coefficient and molar mass of gas |
| $\underline{\pi}^{GW} / \overline{\pi}^{GW} / \underline{q}^{GW} / \overline{q}^{GW}$ | Lower/upper bounds of pressures and mass flow rates of gas wells |
| $\underline{\pi}^{CP} / \overline{\pi}^{CP}$ | Lower/upper bounds of boosted pressures of the compressors |
| $N_{gw} / N_{cp} / N_{gl}$ $/ N_{gf} / N_L$ | Number of gas wells, compressors, traditional gas loads, GFUs, and potential pipeline leakage conditions in the NGS. |

### C. Variables

| | |
|---|---|
| $\xi_t$ | Uncertain power mismatch in the EPS at time $t$ |
| $P_{i,t}$ | Base output of unit $i$ at time $t$ |
| $R_{i,t}^{up} / R_{i,t}^{dw}$ | Upward/downward reserves of unit $i$ at time $t$ |
| $\chi_{i,t}$ | Participation factor of AGC unit $i$ at time $t$ |

## I. Introduction

THE decrease in fossil energy storage and the need for environmental protection have made the development of renewable energy sources (RESs) a consensus in the global energy industry [1]. However, the intermittency of RESs with increasing installed capacity poses challenges to the stable operation of electric power systems (EPSs). Gas-fired units (GFUs) with fast regulation are considered an effective way to mitigate RES fluctuations and have been rapidly developed in recent years, leading to stronger couplings of EPSs and natural gas systems (NGSs) [2]. Therefore, increasing attention has been given to the real-time collaborative dispatch of integrated electric and gas systems (IEGSs) with RES uncertainty.

Benefiting from the tight coupling between the EPS and the NGS, during real-time dispatch, the NGS can provide flexibility for GFUs to balance the EPS power mismatches caused by the generation fluctuations of RESs [2]. However, this tight coupling can also cause local disturbances to propagate to the entire IEGS through GFUs [3], which poses challenges for the real-time dispatch of IEGSs. Since GFUs deal with uncertain RES fluctuations, the uncertainty in the EPS can propagate to the NGS through GFUs, resulting in pressure and gas flow fluctuations in pipeline networks, possibly exceeding the allowable range [4].

This work was supported by the National Natural Science Foundation of China (No. 52377107) and the Shanxi Energy Internet Research Institute (SXEI2023ZD002). *(Corresponding author: Zhengshuo Li.)*

Peiyao Zhao and Zhengshuo Li (e-mail: zsli@sdu.edu.cn) are with the School of Electrical Engineering, Shandong University, 250061, Jinan, China. Jiahui Zhang, Xiang Bai, and Jia Su are with the Shanxi Energy Internet Research Institute, 030002, Shanxi, China.

The real-time economic dispatch (RTED) model for IEGSs has been studied in several works, e.g., [5]-[9]. Ref. [5] highlighted the uncertainty propagation mechanism of a coupling device balancing active power disturbances. Ref. [7]-[9] noted that to preserve data privacy, decentralized solutions are more suitable for the collaborative optimization of an IEGS than centralized solutions, but in practice, the number of information iterations between the EPS and the NGS should be reduced as much as possible. In one word, the RTED model for IEGSs has relatively high requirements on computational time and system operational security [10].

However, it should be noticed that most of the current literature only consider NGS operational constraints under ideal conditions, that is, they rarely consider the dynamic security of NGSs with pipeline failures: *outage* and/or *leakage*. In physics, these pipeline failures may threaten the electricity supply reliability of the EPS through GFUs [11]. Moreover, compared with pipeline outage contingencies that are considered in [12],[13], *pipeline leakage* is more common in NGSs and the leaky pipelines usually cannot be removed immediately [14]. Due to the slow dynamic physics of the NGS [15], the pipeline leakage will result in a failure operation state over a longer time scale, e.g., several hours [16]. In this case, the capacity of an NGS to supply gas to GFUs may continue to be affected, preventing GFUs from implementing real-time dispatch instructions that are optimized under normal conditions.

Currently, there are very few real-time dispatch studies for IEGSs considering pipeline leakages. Ref. [17] proposed a two-stage dispatch model considering pipeline leakage and linepack. However, this study focused on the dispatch under specific initial leakage conditions rather than considering all potential leakage failures. Ref. [16] proposed a risk-based robust dispatch model considering multiple possible pipeline leakage failures, but did not consider the uncertainty propagation between the EPS and NGS. The location and number of leaking pipes in real-time dispatch can be random, henceforth called *pipeline leakage uncertainty*. To the best of our knowledge, there is a lack of an RTED model that considers both the *uncertainty propagation* and *pipeline leakage uncertainty* issues. With the tight coupling between EPSs and NGSs, an RTED dispatch scheme that ignores these two issues may cause serious pressure violations in the NGS, affect the gas supply to GFUs, and weaken the generation of GFUs, thereby threatening EPS operational security.

Nevertheless, considering the two issues in the RTED will bring in the following challenges in the aspects of *modeling* and *solution* methods:

1) To ensure the security of RTED decisions for IEGSs, one should consider the operational constraints regarding all potential pipeline leakage failures, which impose a substantial computational burden on the RTED model. In addition, to characterize the dynamic process of an NGS under pipeline leakage failure, one should use a set of partial differential equations (PDEs), which further increases the complexity of the RTED model [18].

2) Due to widespread concerns about data privacy, the centralized RTED model for IEGSs can be challenging to apply in practice [19]. In addition, since considering RES uncertainty propagation and pipeline leakage uncertainty will introduce large-scale constraints and uncertain state variables into the NGS-side model, conventional decentralized solutions, e.g., the alternating direction method of multipliers (ADMM), may converge slowly due to the excessive number of iterations between subproblems in the EPS and NGS [20].

To address these challenges, this paper first presents a novel RTED model for IEGSs that addresses the issues of uncertainty propagation and uncertain gas pipeline leakage failures, and then provides an efficient decentralized solution based on the *coupling boundary dynamic adjustment region* of the NGS. The main contributions are summarized below.

1) A novel RTED model for IEGSs is proposed that considers the combined effect of RES uncertainty and pipeline leakage uncertainty. The RES uncertainty is characterized by a continuous robust uncertainty model based on sample paths, and the dynamic state changes of the pipeline network under uncertain leakage conditions are characterized by PDEs and a discrete set that includes all potential pipe leakage conditions.

2) A novel notion and formulation are proposed to characterize the feasible range of gas that an NGS can supply to GFUs under arbitrary gas pipeline leakage failures, namely, *the coupling boundary dynamic adjustment region considering pipeline leakage failures* (LCBDAR). This LCBDAR can be seen as *surrogate* constraints in place of the original large-scale and complicated PDE-constrained NGS operation constraints under uncertainty. Based on the LCBDAR, a decentralized solution is established to solve the above RTED model in a *noniterative* mode, in which *noniterative* means that the decisions are optimized with only *one-time information exchange* from the NGS to the EPS. It should be noted that the concept of the noniterative decentralized solution has been explored in [2],[21]; our novelty mainly lies in the proposal and formulation of LCBDAR, which is the key to developing such a noniterative decentralized solution.

3) Due to the large number of potential leakage failures, assessing the LCBDAR is still a large-scale and complicated problem. By leveraging the latent decomposable structure of this problem, we further propose an efficient and parallel-implementable algorithm to accurately assess the LCBDAR in a much-shortened time, and as a result, to significantly reduce the total computational time of the aforesaid noniterative decentralized solution. In our case studies, the speedup ratio of our decentralized solution over the common algorithms, e.g., ADMM, can be as high as 9-fold.

The rest of the paper is organized as follows. Section II introduces the coupling between an EPS and an NGS. Section III develops the NGS dynamic model considering pipeline leakage uncertainty and the RTED model for IEGS. Section IV presents the LCBDAR and the customized algorithms for assessing it. Section V introduces the decentralized framework based on LCBDAR. The case studies and results are described in Section VI. Conclusions are drawn in Section VII.

## II. COUPLING BETWEEN THE EPS AND NGS

The configuration of an IEGS is sketched in Fig. 1. During real-time dispatch, GFUs participate in automatic generation control (AGC) to balance the real-time power mismatches caused by uncertain wind generation fluctuations [22]; therefore, the generation outputs and the corresponding gas withdrawals of GFUs are time-varying.

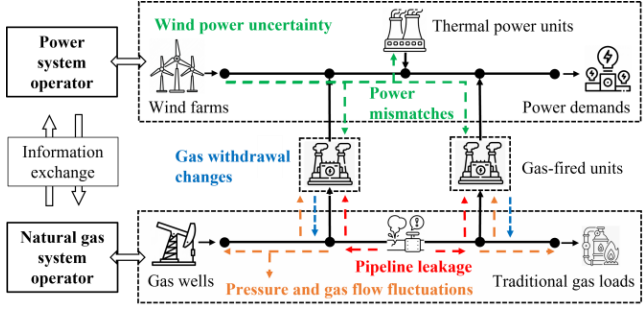

Fig. 1. A typical structure of an IEGS

Due to the NGS dynamic characteristics, the uncertain gas withdrawal changes of GFUs will lead to continuous fluctuations in the gas flow rates and pressures in the gas pipeline network. In particular, when the power mismatches in the EPS are noticeable, the corresponding gas withdrawal changes of GFUs may result in significant fluctuations in gas flow rates and pressures in the natural gas pipelines, and there will be a risk of pipeline pressure violations and other security issues during the real-time operation of the NGS.

The aforementioned security issues caused by insecure gas withdrawals of GFUs, as well as pipeline leakage failures, can further affect the normal gas supply of an NGS to GFUs and weaken the flexible regulation capability of GFUs to address power mismatches in the EPS, which will ultimately threaten the operational stability of the EPS. Therefore, these security issues should be considered in the RTED to ensure the real-time operational security of the whole IEGS.

### III. THE RTED MODEL CONSIDERING THE UNCERTAINTY PROPAGATION AND PIPELINE LEAKAGE UNCERTAINTY

RTED is usually executed every 15 minutes to determine the generation schedules in the next 1 hour, with a temporal resolution of 5 minutes [23]. In what follows, we denote the time window spanned by each RTED as $T$, the temporal resolution as $\tau$, and the number of time intervals in one RTED time window as $N_\tau = T/\tau$.

*A. EPS-Side Model with RES Uncertainty*

The real-time generation of a wind farm can be modeled as the sum of the forecast generation output $P_{i,t}^w$ and the uncertain forecast deviation $\zeta_{i,t}$:

$$p_{i,t}^w = P_{i,t}^w + \zeta_{i,t}, \quad \forall t \in N_\tau \ \forall i \in \Omega_W \quad (1)$$

The systemwide power mismatch $\xi_t$ in the EPS can be considered approximately equal to the sum of the forecast deviations $\zeta_{i,t}$ of all the wind power farms, i.e., $\xi_t \approx \sum_{i \in \Omega_W} \zeta_{i,t}$.

Inspired by [24], a *robust sample path-based uncertainty model* for $\xi_t$ is established in this paper. The reason for adopting this model is that it does not require any parametric assumptions regarding the probability distribution of uncertain variables, and enables the RTED model to be reformulated as a computationally tractable linear robust optimization with satisfactory out-of-sample performance.

*Remark 1*: In common uncertainty models, probabilistic scenario-based models for stochastic optimization depend excessively on the wind probability distribution, which is difficult to estimate accurately in practice [25]; traditional uncertainty sets for robust optimization usually lead to relatively conservative decisions [26]; ambiguity sets used in distributionally robust optimization may make optimization models intractable and need to be converted into approximate solvable forms that are usually large-scale and mathematically complicated [27].

In this *robust sample path-based uncertainty* model, the subuncertainty sets $\mathcal{U}_N^\ell$ based on sample paths are established for the uncertain variable $\xi_t$:

$$\mathcal{U}_N^\ell = \left\{ \xi_t \in \mathbb{R}, \forall t \in N_\tau : \partial_t^\ell \leq \xi_t \leq \vartheta_t^\ell \right\}, \ell \in \mathcal{N} \quad (2)$$

where $\mathcal{N}$ denotes the sample path set of historical data of $\boldsymbol{\xi} = \{\xi_t\}_{t \in N_\tau}$. $\ell$ denotes the index of sample paths. $\vartheta_t^\ell$ and $\partial_t^\ell$ are the upper and lower bounds that define the uncertainty set. The complete robust uncertainty model based on the sample paths of $\xi_t$, denoted by $\mathcal{U}$, can be expressed as the union of subuncertainty sets $\mathcal{U}_N^\ell$, i.e., $\mathcal{U} = \bigcup_{\ell=1}^N \mathcal{U}_N^\ell$.

*Remark 2*: $\mathcal{U}_N^\ell$ can be considered as relaxing a sample path into a prediction interval considering more potential scenarios, which enhances the robustness of the stochastic RTED. Moreover, $\mathcal{U}_N^\ell$ is only built around the sample path to avoid low-probability or impossible scenarios, which reduces the conservativeness of RTED decisions.

Moreover, according to the physical characteristics of the AGC methods often adopted in the literature on EPS dispatch, e.g., [28],[29], the AGC regulation process can be formulated as the following linear decision rule:

$$\tilde{P}_{i,t} = P_{i,t} + \chi_{i,t}\xi_t, \ \sum_{i \in \Omega_A} \chi_{i,t} = 1, \ \chi_{i,t} \geq 0, \ \forall i \in \Omega_A \ \forall t \in N_\tau \quad (3)$$

where $\tilde{P}_{i,t}$ denotes the real-time output of unit $i$ at time interval $t$, $P_{i,t}$ denotes the base output, $\chi_{i,t}$ denotes the AGC participation factor, $\chi_{i,t}\xi_t$ denotes the AGC regulation output.

The objective of the RTED model is to minimize the cost of the base outputs $P_{i,t}$, the upward/downward reserves $R_{i,t}^{up}/R_{i,t}^{dw}$, and the average cost of the AGC regulation outputs of units over all the uncertainty sets $\mathcal{U}_N^\ell$ [30]:

$$\min_{x} \left\{ \sum_{t=1}^{N_\tau} \sum_{i \in \Omega_A} \left[ c_i^u P_{i,t} + c_i^{up} R_{i,t}^{up} + c_i^{dw} R_{i,t}^{dw} \right] + \frac{1}{N} \sum_{n=1}^{N} \sup_{\xi_t \in \mathcal{U}_N^\ell} \sum_{t=1}^{N_\tau} \sum_{i \in \Omega_A} \left[ c_i^r \chi_{i,t} \xi_t \right] \right\} \quad (4a)$$

The constraints of the RTED model include power balance constraints (4b), system reserve requirements (4c), limitations of transmission lines (4d), limitations of generator capacity for units (4e), limitations of ramping capability for units (4f) and limitations of regulated generation output for AGC units (4g):

$$\sum_{i \in \Omega_A} P_{i,t} + \sum_{i \in \Omega_W} P_{i,t}^w = \sum_{i \in \Omega_L} D_{i,t} \quad \forall t \in N_\tau \quad (4b)$$

$$\sum_{i \in \Omega_A} R_{i,t}^{up} \geq R_{sys}^{up}, \sum_{i \in \Omega_A} R_{i,t}^{dw} \geq R_{sys}^{dw}, \ \forall t \in N_\tau \quad (4c)$$

$$\begin{cases} -L_m^{tr} \leq \sum_{s \in \Omega_S} H_{m,s}^{PT} \left[ \sum_{i \in \Omega_{U,s}} \tilde{P}_{i,t} + \sum_{i \in \Omega_{W,s}} p_{i,t}^w - \sum_{i \in \Omega_{L,s}} D_{i,t} \right] \leq L_m^{tr}, \forall t \in N_\tau \\ -L_m^{tr} \leq \sum_{s \in \Omega_S} H_{m,s}^{PT} \left[ \sum_{i \in \Omega_{U,s}} P_{i,t} + \sum_{i \in \Omega_{W,s}} P_{i,t}^w - \sum_{i \in \Omega_{L,s}} D_{i,t} \right] \leq L_m^{tr}, \forall t \in N_\tau \end{cases} \quad (4d)$$

$$P_{i,t} + R_{i,t}^{up} \leq P_i^{\max}, \ P_{i,t} - R_{i,t}^{dw} \geq P_i^{\min}, \ \forall i \in \Omega_A \ \forall t \in N_\tau \quad (4e)$$

$$\tilde{P}_{i,t} - \tilde{P}_{i,t-1} \leq r_i^{up}\tau, \ \tilde{P}_{i,t-1} - \tilde{P}_{i,t} \leq r_i^{dw}\tau, \ \forall i \in \Omega_A \ \forall t \in N_\tau \quad (4f)$$

$$\chi_{i,t}\xi_t \leq R_{i,t}^{up}, \ -\chi_{i,t}\xi_t \leq R_{i,t}^{dw}, \ \forall i \in \Omega_A \ \forall t \in N_\tau \ \forall \xi_t \in \mathcal{U} \quad (4g)$$

To facilitate the next derivation, the RTED model is summarized into the following compact matrix-vector form:

$$\min_{\boldsymbol{x}_t^b, \boldsymbol{x}_t^s} \left\{ \sum_{t=1}^{N_\tau} \boldsymbol{c}_b^{\mathrm{T}} \boldsymbol{x}_t^b + \frac{1}{N} \sum_{n=1}^{N} \sup_{\xi_t \in \mathcal{U}_N^\ell} \sum_{t=1}^{N_\tau} \boldsymbol{c}_s^{\mathrm{T}} \boldsymbol{x}_t^s \xi_t \right\} \quad (5a)$$

$$\text{s.t.} \begin{cases} \max_{\xi_t \in \mathcal{U}_N^\ell} \left( \boldsymbol{A}_t \boldsymbol{x}_t^s \xi_t + \boldsymbol{B}_t \boldsymbol{x}_{t-1}^s \xi_t \right) + \boldsymbol{C}_t \boldsymbol{x}_t^b + \boldsymbol{D} \boldsymbol{x}_{t-1}^b \leq \boldsymbol{a}_t, \; \forall t \in N_\tau \\ \boldsymbol{E}_t \boldsymbol{x}_t^s + \boldsymbol{F}_t \boldsymbol{x}_t^b = \boldsymbol{b}_t, \; \forall t \in N_\tau \end{cases} \quad (5b)$$

where the decision variables include $\boldsymbol{x}_t^b := \{P_{i,t}, R_{i,t}^{up}, R_{i,t}^{dw}\}_{i \in \Omega_A}^{\mathrm{T}}$ and $\boldsymbol{x}_t^s := \{\chi_{i,t}\}_{i \in \Omega_A}^{\mathrm{T}}$ at $N_\tau$ time intervals.

*B. NGS-Side Dynamic Security Model Considering Uncertain Pipeline Leakage*

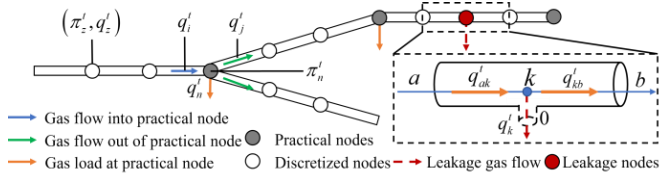

Fig. 2. Discretized gas pipeline network with pipeline leakage.

The dynamic changes in the pressure $\pi$ and mass flow rate $q$ in a pipeline can be described by a set of PDEs, which are provided in the electronic companion [31]. To facilitate the subsequent optimization procedure, the pipeline network with leakage failures is discretized as shown in Fig. 2, and the continuous pressure and flow distributions in pipelines are discretized into pressure and mass flow rate variables at practical nodes, discretized nodes, and leakage nodes. Thus, the PDEs can be discretized with the central difference method [32] and linearized with the average flow velocity (AFV) [15]:

$$\begin{cases} \left( \dfrac{\pi_{z+1}^{t+1} - \pi_{z+1}^t + \pi_z^{t+1} - \pi_z^t}{\Delta t} \right) + \dfrac{c^2}{S} \left( \dfrac{q_{z+1}^{t+1} - q_z^{t+1} + q_{z+1}^t - q_z^t}{\Delta z} \right) = 0, \\ \left( \dfrac{\pi_{z+1}^{t+1} - \pi_z^{t+1} + \pi_{z+1}^t - \pi_z^t}{\Delta z} \right) + \dfrac{1}{S} \left( \dfrac{q_{z+1}^{t+1} - q_{z+1}^t + q_z^{t+1} - q_z^t}{\Delta t} \right) \\ + \dfrac{\lambda \bar{v}_z^t}{4dS} \left( q_{z+1}^{t+1} + q_{z+1}^t + q_z^{t+1} + q_z^t \right) = 0, \forall t \in N_\tau, \; \forall z \in \Omega_z^w, \; \forall w \in \mathcal{W} \end{cases} \quad (6)$$

where $z$ denotes the index of nodes. $\bar{v}_z^t$ denotes the AFV of gas flow. $\Omega_z^w$ denotes the set of nodes with discretized state variables under pipeline leakage condition $w$. $\mathcal{W}$ denotes the discrete set that includes the normal condition and all potential pipeline leakage conditions of the pipeline network.

For a pipeline leakage at node $k$ in pipe segment $ab$, as shown in Fig. 2, the gas flow of the leakage hole can be considered isentropic flow, and the leakage hole is usually smaller than the cross-sectional area of the whole pipeline [33]. Thus, the critical flow model is used to describe the dynamic leakage process, and the leakage flow equation can be written as follows [34]:

$$q_k^t = \mu_k A_k \pi_k^t \sqrt{\dfrac{\kappa M}{c^2} \left( \dfrac{2}{\kappa+1} \right)^{(\kappa+1)/(\kappa-1)}}, \forall t \in N_\tau \; \forall k \in \Omega_k^w \; \forall w \in \mathcal{W} \quad (7)$$

where $q_k^t$ and $\pi_k^t$ denote the mass flow rate and pressure in leakage node $k$ at time $t$, respectively, and $\Omega_k^w$ denotes the set of pipeline leakage nodes under leakage condition $w$.

Moreover, the pressure continuity and flow balance at the junctions of pipelines and leakage nodes should also be satisfied and are formulated as follows:

$$\begin{cases} \pi_i^t = \pi_n^t, \; \forall t \in N_\tau, \; \forall i \in \Omega_n \\ \sum_{i \in \Omega_n^I} q_i^t - \sum_{j \in \Omega_n^O} q_j^t - q_n^t = 0, \; \forall t \in N_\tau \\ \pi_{ak}^t = \pi_k^t = \pi_{kb}^t, \; \forall t \in N_\tau, \; \forall k \in \Omega_k^w, \; \forall w \in \mathcal{W} \\ q_{ak}^t - q_{kb}^t - q_k^t = 0, \; \forall t \in N_\tau, \; \forall k \in \Omega_k^w, \; \forall w \in \mathcal{W} \end{cases} \quad (8)$$

where $\Omega_n$ denotes the set of pressures $\pi_i^t$ at the end of the pipe connected to node $n$; $\Omega_n^I$ and $\Omega_n^O$ denote the sets of gas flows into and out of practical node $n$, respectively. $q_n^t$ denotes the gas load at node $n$. $\pi_{ak}^t$ and $\pi_{kb}^t$ denote the pressures at the ends of pipe segments $ak$ and $kb$ connected to the leakage node $k$, respectively. $q_{ak}^t$ and $q_{kb}^t$ denote the mass flow rates in pipe segments $ak$ and $kb$, respectively.

The complete NGS dynamic state model under condition $w$ can be summarized by (6)-(8). The NGS state vector at time $t$ is denoted by $\boldsymbol{y}_t(w)$, which includes the pressures and mass flow rates at discretized and leakage nodes. The gas well pressure vector is denoted by $\boldsymbol{\pi}_t^{GW} := \{\pi_{i,t}^{GW}\}_{i \in \Omega_{GW}}^{\mathrm{T}}$, the boosted pressure vector of compressors is denoted by $\boldsymbol{\pi}_t^{CP} := \{\pi_{i,t}^{CP}\}_{i \in \Omega_{CP}}^{\mathrm{T}}$, the traditional gas load vector is denoted by $\boldsymbol{q}_t^{GL} := \{q_{i,t}^{GL}\}_{i \in \Omega_{GL}}^{\mathrm{T}}$, and the GFU gas withdrawal vector is denoted by $\boldsymbol{q}_t^{GF} := \{q_{i,t}^{GF}\}_{i \in \Omega_G}^{\mathrm{T}}$. The NGS dynamic state model (6)-(8) can be rewritten as:

$$\begin{bmatrix} \boldsymbol{M}_t^{PDE}(w) \\ \boldsymbol{M}_t^B(w) \end{bmatrix} \begin{bmatrix} \boldsymbol{\pi}_t^{GW}(w) \\ \boldsymbol{y}_t(w) \\ \boldsymbol{\pi}_t^{CP}(w) \\ \boldsymbol{q}_t^{GL}(w) \\ \boldsymbol{q}_t^{GF}(w) \end{bmatrix} = \boldsymbol{M}_{t-1}^{PDE}(w) \begin{bmatrix} \boldsymbol{\pi}_{t-1}^{GW}(w) \\ \boldsymbol{y}_{t-1}(w) \end{bmatrix} \quad (9)$$

where $\boldsymbol{M}_t^{PDE}$ denotes the coefficient matrix for (6), and $\boldsymbol{M}_t^B$ denotes the coefficient matrix for (7)-(8). Since (7)-(8) do not include the variables at time $t$-1, the corresponding terms on the right side of (9) are equal to zero. Below the matrix elements are the dimensions, where $N_p$ and $N_{pf}$ denote the numbers of equations (6) and (7)-(8) for the entire NGS, respectively. $N_d$ and $N_k$ denote the numbers of discretized nodes and leakage nodes, respectively.

According to (9), the NGS state $\boldsymbol{y}_t$ and the corresponding contents of the coefficient matrix $\boldsymbol{M}_t^{PDE}, \boldsymbol{M}_t^B$ are directly associated with the leakage condition $w$, and the values of the variables $\boldsymbol{\pi}_t^{GW}, \boldsymbol{\pi}_t^{CP}, \boldsymbol{q}_t^{GL}, \boldsymbol{q}_t^{GF}$ will also be affected by the leakage condition $w$, i.e., the NGS states and variables under different leakage conditions can differ. In addition, the overall format of the NGS dynamic state model under different leakage conditions is consistent, which means that the model (9) under different pipeline conditions can be reformulated as a unified compact matrix–vector formula:

$$\begin{aligned} \boldsymbol{y}_t(w) = {}& \boldsymbol{S}(w) \boldsymbol{y}_{t-1}(w) + \boldsymbol{s}(w) \boldsymbol{\pi}_{t-1}^{GW}(w) + \boldsymbol{I}_1(w) \boldsymbol{\pi}_t^{GW}(w) \\ & + \boldsymbol{I}_2(w) \boldsymbol{\pi}_t^{CP}(w) + \boldsymbol{I}_3(w) \boldsymbol{q}_t^{GL}(w) + \boldsymbol{I}_4(w) \boldsymbol{q}_t^{GF}(w), \quad (10) \\ & \forall t \in N_\tau \; \forall w \in \mathcal{W} \end{aligned}$$

where $S, s, I_1 \sim I_4$ are the coefficient matrices converted from $M_t^{PDE}, M_t^B$ in (9), which depend on the leakage condition $w$.

Note that gas load shedding under pipeline leakage conditions is considered in this paper, and the traditional gas load $q_t^{GL}$ satisfies the following constraints:

$$q_t^{GL}(w) = q_t^{FL} - q_t^S(w), \; t \in N_\tau \; \forall w \in \mathcal{W} \quad (11)$$

where $q_t^{FL}$ denotes the forecasted traditional gas load and $q_t^S$ denotes the controllable load shedding.

The pressures of nodes, gas wells, and compressors, as well as the gas flows of gas wells, should stay within the predefined secure range during the above dynamic process [2]:

$$\begin{cases} H_t(w) y_t(w) \le g_t(w), & \forall t \in N_\tau \; \forall w \in \mathcal{W} \\ \underline{\pi}^{GW} \le \pi_t^{GW}(w) \le \overline{\pi}^{GW}, & \forall t \in N_\tau \; \forall w \in \mathcal{W} \\ \underline{q}^{GW} \le q_t^{GW}(w) \le \overline{q}^{GW}, & \forall t \in N_\tau \; \forall w \in \mathcal{W} \\ \underline{\pi}^{CP} \le \pi_t^{CP}(w) \le \overline{\pi}^{CP}, & \forall t \in N_\tau \; \forall w \in \mathcal{W} \end{cases} \quad (12)$$

where $H_t$ and $g_t$ are coefficient matrices and $q_t^{GW}$ denotes the vector of gas flows of gas wells. The complete dynamic security model for an NGS can be summarized as a dynamic state model (10) and a set of security constraints (12).

According to the matrix and variable dimensions in (9), after considering the time dimension, the scales of the NGS dynamic security model under each leakage condition can be even larger; when considering the dynamic security of the NGS with the pipeline leakage uncertainty, one should incorporate the constraints under all potential pipeline leakage conditions in the RTED model, which will significantly increase the complexity of the analysis and calculation.

### C. Coupling Constraints and Uncertainty Propagation

GFUs serve as coupling devices for the EPS and NGS. The relationship between the gas withdrawals and the real-time generation outputs of GFUs can be described as follows:

$$\begin{cases} q_{i,t}^{GF}(\xi, w) = c_2^G \left(\tilde{P}_{i,t}\right)^2 + c_1^G \left(\tilde{P}_{i,t}\right) + c_0^G, \forall t \in N_\tau \; \forall i \in \Omega_G \; \forall w \in \mathcal{W} \\ \tilde{P}_{i,t} = P_{i,t} + \chi_{i,t} \xi_t, \forall t \in N_\tau \; \forall i \in \Omega_G \; \forall \xi_t \in \mathcal{U} \end{cases} \quad (13)$$

where $c_2^G, c_1^G, c_0^G$ are coefficients of the gas consumption characteristics of GFUs.

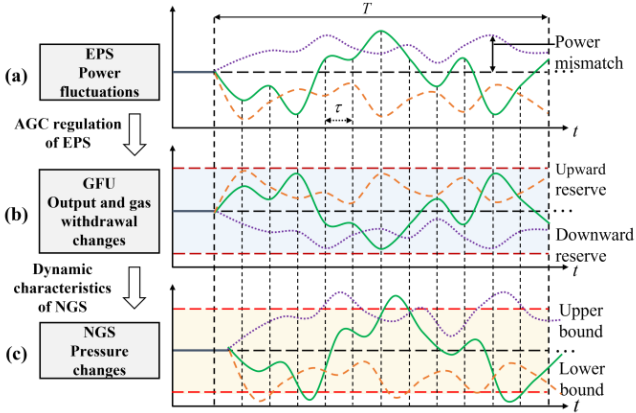

Fig. 3. The propagation of uncertainty from an EPS to an NGS

However, GFUs coupling the EPS and NGS can also lead to uncertainty propagation, illustrated by Fig. 3. In the context of GFUs participating in AGC to balance the power fluctuations in the EPS, the gas withdrawals of GFUs are conditioned by the uncertain power mismatch $\xi_t$ according to (13). Following the NGS dynamic state model (10), $y_t^k$ is thus affected by $\xi_t$, which means that the uncertainties in the EPS propagate to the NGS; as a result, the states and control in the whole IEGS will be affected by the uncertain variable $\xi_t$.

### D. Complete RTED Model for IEGSs

By indicating the dependence of the variables on the uncertain parameters, the complete RTED model considering RES uncertainty propagation and uncertain gas pipeline leakage failures is formulated as follows:

$$\text{Obj} \min_{x_t^b, x_t^s} \left\{ \sum_{t=1}^{N_\tau} c_b^T x_t^b + \frac{1}{N} \sum_{n=1}^{N} \sup_{\xi_t \in \mathcal{U}_N^t} \sum_{t=1}^{N_\tau} c_s^T x_t^s \xi_t \right\}$$

s.t. $\forall t \in N_\tau, \; \forall (\xi_t, w) \in \mathcal{U} \times \mathcal{W}$

$$\text{EPS} \begin{cases} \max_{\xi_t \in \mathcal{U}_N^t} \left( A_t x_t^s \xi_t + B_t x_{t-1}^s \xi_t \right) + C_t x_t^b + D x_{t-1}^b \le a_t \\ E x_t^s + F_t x_t^b = b_t \end{cases} \quad (14)$$

$$\text{NGS} \begin{cases} y_t(\xi, w) = S(w) y_{t-1}(w) + s(w) \pi_{t-1}^{GW}(w) + I_1(w) \pi_t^{GW}(w) \\ \qquad + I_2(w) \pi_t^{CP}(w) + I_3(w) q_t^{GL}(w) + I_4(w) q_t^{GF}(\xi, w) \\ H_t(w) y_t(\xi, w) \le g_t(w), \; \underline{\pi}^{CP} \le \pi_t^{CP}(w) \le \overline{\pi}^{CP}, \\ \underline{\pi}^{GW} \le \pi_t^{GW}(w) \le \overline{\pi}^{GW}, \; \underline{q}^{GW} \le q_t^{GW}(w) \le \overline{q}^{GW} \end{cases}$$

$$\text{Coupling} \begin{cases} q_t^{GF}(\xi, w) = c_2^G \left[ \tilde{P}_t^{GF}(\xi_t) \circ \tilde{P}_t^{GF}(\xi_t) \right] + c_1^G \left[ \tilde{P}_t^{GF}(\xi_t) \right] + c_0^G \\ \tilde{P}_t^{GF}(\xi_t) = K x_t^b + L x_t^s \xi_t \end{cases}$$

where "$\circ$" denotes the Hadamard product. "$\times$" denotes the Cartesian product. $\tilde{P}_t^{GF} := \{\tilde{P}_{i,t}\}_{i \in \Omega_G}^T$ denotes the vector of the real-time generation outputs of GFUs at time $t$.

In model (14), due to the uncertainty propagation, the dynamic state variables $y_t(\xi, w)$ in the NGS are affected by both the RES uncertainty in the EPS and the pipeline leakage uncertainty in the NGS; as a result, the computational complexity caused by the mixed uncertainty set $\mathcal{U} \times \mathcal{W}$ can be prohibitive. To address this difficulty, a novel LCBDAR and an LCBDAR-based decentralized framework are developed below to efficiently solve the RTED model (14).

## IV. COUPLING BOUNDARY DYNAMIC ADJUSTMENT REGION CONSIDERING PIPELINE LEAKAGE FAILURES

### A. Notion and Formulation of the LCBDAR

The LCBDAR characterizes the ranges of gas that the NGS can securely supply to GFUs under arbitrary pipeline leakage conditions; i.e., the aim is to find a feasible region $\mathcal{R}$ in the vector space of the gas withdrawal $q^{GF} := \{q_t^{GF}\}_{t \in N_\tau}^T$ that is as large as possible. For arbitrary $q^{GF}$ in $\mathcal{R}$, under arbitrary leakage conditions in $\mathcal{W}$, feasible decisions $\pi^{GW} := \{\pi_t^{GW}\}_{t \in N_\tau}^T$ and $\pi^{CP} := \{\pi_t^{CP}\}_{t \in N_\tau}^T$ always exist that ensure that the NGS state $y_t$ resulting from (10) will satisfy the NGS security constraints (12) without gas load shedding $q^S := \{q_t^S\}_{t \in N_\tau}^T$. The exact mathematical definition of the LCBDAR is:

> For $\forall w \in \mathcal{W}$, $\forall q^{GF}(\xi, w) \in \mathcal{R}$, $\exists \pi^{GW}(w)$, $\exists \pi^{CP}(w)$, and $\exists q^S(w) = 0$ such that (10)-(12) hold.

Hence, the exact LCBDAR can be formulated as follows:

$$\mathcal{R} := \left\{ \left( q^{GF}(\xi,w), \pi^{GW}(w), \pi^{CP}(w), q^{S}(w) \right) \forall w \in \mathcal{W} \mid (10)-(12) \right\} \quad (15)$$

However, due to the complex dynamic state relation given by the discretized PDEs and pipeline leakage uncertainty, it is intractable to compute and apply the exact LCBDAR $\mathcal{R}$ (15). Thus, a box-shaped inner approximation of $\mathcal{R}$ is generally adopted [35], denoted by $\mathcal{Q}$, which can be constructed by solving the following problem:

> To find the $\mathcal{Q} := \left[ q^{d}(\mathcal{W}), q^{u}(\mathcal{W}) \right]$ with largest volume, for $\forall w \in \mathcal{W}$, $\forall q^{GF}(\xi,w) \in \mathcal{Q}$, $\exists \pi^{GW}(w)$, $\exists \pi^{CP}(w)$, and $\exists q^{S}(w) = 0$ such that (10)-(12) hold.

where $q^{u}$ and $q^{d}$ denote the vector of upper and lower bounds of gas withdrawal $q^{GF}$, respectively. $q^{u}$ and $q^{d}$ are affected by all the leakage conditions in $\mathcal{W}$. To simplify the expression, with a slight abuse of notation, we omit the symbols indicating the dependence of $q^{u}$ and $q^{d}$ on $\mathcal{W}$ below. Note that the LCBDAR in the following models refers to the maximum inner-approximated LCBDAR $\mathcal{Q}$.

To establish the mathematical form of the above problem of LCBDAR $\mathcal{Q}$, we first introduce an auxiliary uncertainty variable $\boldsymbol{u} := \{\{u_{i,t}\}_{i \in \Omega_G}\}_{t \in N_\tau}^{\mathrm{T}}$ to equivalently reformulate arbitrary $q^{GF}(\xi,w)$ in $\mathcal{Q}$ into the linear combination of $q^{u}$ and $q^{d}$:

$$q^{GF}(w) = \boldsymbol{u} \circ q^{u} + (1-\boldsymbol{u}) \circ q^{d} \quad (16)$$

and $\boldsymbol{u}$ is in the following uncertainty set:

$$\mathbb{U}(\boldsymbol{u}) = \left[ u_{i,t}, \forall t \in N_\tau, \forall i \in \Omega_G \mid 0 \leq u_{i,t} \leq 1 \right] \quad (17)$$

**Remark 3**: The motivation for adopting uncertainty variables $\boldsymbol{u}$ and model (16)-(17) is to characterize the uncertain fluctuation of gas withdrawal $q^{GF}$ caused by GFUs balancing the EPS-side uncertain power mismatch $\xi$, so as to consider the impact of uncertainty propagation on NGS states and decision variables in the following NGS-side model meanwhile avoid introducing EPS-side uncertainty variables.

With the model (16)-(17), similar to the analysis in Ref. [35] and [36], the above problem of $\mathcal{Q}$ can be formulated into the following two-stage adaptive robust optimization model, namely the *LCBDAR assessment model*:

$$\max_{q^{u},q^{d}} \mathbf{1}^{\mathrm{T}}(q^{u}-q^{d}) + \min_{\boldsymbol{u} \in \mathbb{U}} \max_{z} V z(w,\boldsymbol{u}) \quad (18a)$$

s.t. $q^{u} \geq q^{d}$, $q^{u} \geq 0$, $q^{d} \geq 0$,

$$\forall w \in \mathcal{W}, \forall \boldsymbol{u} \in \mathbb{U}, \exists z(w,\boldsymbol{u}) : \begin{cases} R(w) z(w,\boldsymbol{u}) \leq h(w) & (18b) \\ \boldsymbol{u} \circ q^{u} + (1-\boldsymbol{u}) \circ q^{d} = N z(w,\boldsymbol{u}) & (18c) \end{cases}$$

The first-stage variables $q^{u}$ and $q^{d}$ denote the optimization variables that maximize the LCBDAR $\mathcal{Q}$. The second-stage variable $z := (\pi^{GW}, \pi^{CP}, q^{S}, q^{GF})^{\mathrm{T}}$ denotes the adaptive decision that can change the NGS state $y_t$ according to model (10). The minimax optimization objective in (18a) is a penalty item used to minimize the gas load shedding $q^{S}$, where $V$ is the coefficient matrix used to select $q^{S}$ from $z$ and sum $q^{S}$ values.

The linear constraint (18b) is converted from (10)-(12); to reduce the computational burden, the large-scale state $y_t$ is eliminated by adopting the recursion given in (10) and substituting it into (12), $R$ and $h$ are the coefficient matrices.

The constraint (18c) is equivalent to (16), where $N$ is the coefficient matrix used to select $q^{GF}$ from $z$.

The well-known column-and-constraint generation (C&CG) algorithm [37] is generally used for solving two-stage robust optimization problems. However, there are still two types of uncertainties in model (18), i.e., the continuous uncertain gas withdrawal fluctuations of GFUs $\forall \boldsymbol{u} \in \mathbb{U}$ and the discrete uncertain pipeline leakage conditions $\forall w \in \mathcal{W}$, in particular, the latter introduces $T \times N_L \times (N_{gw} + N_{cp} + N_{gl} + 2N_{gf})$ variables and at least $T \times N_L \times (2N_{gw} + N_{cp} + 2N_d + 2N_k)$ constraints, which make it computationally expensive for traditional C&CG to solve model (18). To address this difficulty, an improved parallel column-and-constraint generation (*PC&CG*) algorithm with optimality and convergence guarantees is proposed below to solve model (18) efficiently.

*B. Improved Parallel C&CG Algorithm*

By observing model (17), it is found that the adaptive decision $z(w,\boldsymbol{u})$ has the following two desired properties:

1) $\mathcal{W}$ is a discrete set, and the adaptive decisions $z(w,\boldsymbol{u})$ under different leakage conditions in $\mathcal{W}$ can be enumerated as $\left[ z^{1}(\boldsymbol{u}^{1}), z^{2}(\boldsymbol{u}^{2}) \ldots z^{w}(\boldsymbol{u}^{w}) \right]$ unconditionally.

2) When the outer variables $q^{u}$ and $q^{d}$ are given, the adaptability makes $z^{w}(\boldsymbol{u}^{w})$ independent under different gas withdrawal fluctuation scenarios and leakage conditions.

Based on the above properties, the following equivalent conditions can be given:

$$\begin{aligned} &\forall w \in \mathcal{W}, \forall \boldsymbol{u} \in \mathbb{U}, \exists z(w,\boldsymbol{u}) : \begin{cases} R(w) z(w,\boldsymbol{u}) \leq h(w) \\ \boldsymbol{u} \circ q^{u} + (1-\boldsymbol{u}) \circ q^{d} = N z(w,\boldsymbol{u}) \end{cases} \\ \Leftrightarrow \quad &\text{For every } w \text{ in } \mathcal{W}, \forall \boldsymbol{u}^{w} \in \mathbb{U}, \quad \begin{cases} R^{w} z^{w}(\boldsymbol{u}) \leq h^{w} \\ \boldsymbol{u}^{w} \circ q^{u} + (1-\boldsymbol{u}^{w}) \circ q^{d} = N z^{w}(\boldsymbol{u}) \end{cases} \end{aligned} \quad (19)$$

The equivalent model in (19) can be further solved using the proposed PC&CG algorithm, the flow chart of which is summarized in Fig. 4, where the master problem is the deterministic linear optimization problem shown in (20) and the parallel subproblems are minimax optimizations with uncertainty variables, as shown in (21), which can be solved by adopting strong duality and big-M linearization. The detailed solution process of the PC&CG algorithm is provided in the electronic companion to this paper [31].

$$\text{Obj}: f_M = \max_{q^{u},q^{d}} \mathbf{1}^{\mathrm{T}}(q^{u}-q^{d})$$

s.t. $q^{u} \geq q^{d}$, $q^{u} \geq 0$, $q^{d} \geq 0$, $\quad (20)$

$$\forall i \in [1,\ldots I], \forall w \in \mathcal{W}, \exists z_i^{w} : \begin{cases} R^{w} z_i^{w} \leq h^{w} \\ \boldsymbol{u}_i^{w,*} \circ q^{u} + (1-\boldsymbol{u}_i^{w,*}) \circ q^{d} = N z_i^{w} \end{cases}$$

$$\text{Obj}: f_S^{w} = \min_{\boldsymbol{u}^{w} \in \mathbb{U}} \max_{z^{w}} V z^{w}$$

s.t. $\boldsymbol{u}^{w} \circ q_i^{u,*} + (1-\boldsymbol{u}^{w}) \circ q_i^{d,*} = N z^{w}$

$\quad R^{w} z^{w} \leq h^{w} \quad (21)$

$\quad z^{w} \geq 0, \forall w \in \mathcal{W}, \forall \boldsymbol{u}^{w} \in \mathbb{U}$

where $i$ denotes the index of the iteration and "*" denotes the given value of the corresponding variable. The superscript "$w$" indicates the corresponding coefficient matrices or variables under leakage condition $w$.

In each iteration, the master problem (20) optimizes the upper and lower bounds of LCBDAR $\mathcal{Q}$ and sends it to the subproblem under each pipeline network condition in $\mathcal{W}$. The subproblems (21) are solved in parallel to find the worst gas withdrawal scenarios $\boldsymbol{u}_i^w$ for the GFUs, in which there is no feasible adaptive decision $\boldsymbol{z}^w$ that can satisfy NGS security constraints without any gas load shedding. Then, the new constraints under the worst-case scenario $\boldsymbol{u}_i^{w,*}$ are generated and added to the master problem to improve the optimal solution. In practice, with a tolerance $\delta$ close to zero, iteration can be terminated the first time gas load shedding $f_S < \delta$ occurs, which means that the current LCBDAR $\mathcal{Q}$ is already robust and feasible.

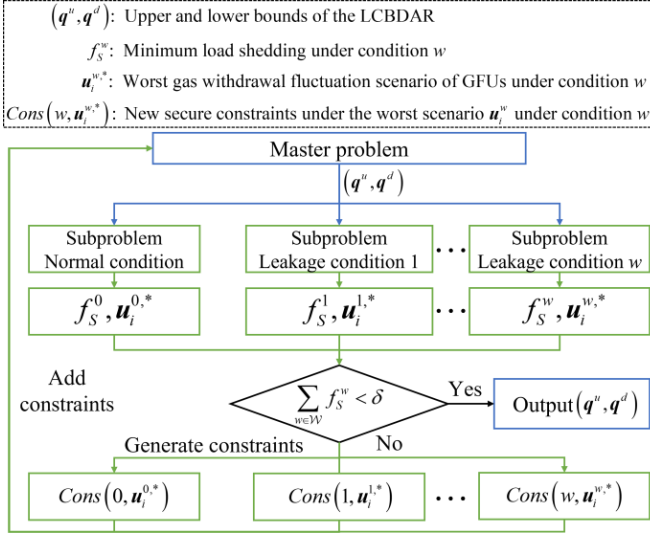

Fig. 4. Flow chart of the PC&CG algorithm.

***Optimality and convergence guarantee***: The proposed PC&CG algorithm has the following properties:

1) The parallel subproblems (21) are all linear programming (LP) problems. The parallel solution simply increases the number of linear constraints with the recourse decision variables of the master problem, and the relatively complete recourse assumption still holds, which means that these LPs are feasible for given $\boldsymbol{q}^u, \boldsymbol{q}^d$ and $\boldsymbol{u}$. This makes the convergence of PC&CG essentially equivalent to that of C&CG.

2) With the disjoint vector space of the uncertain variable $\boldsymbol{u}$ and decision $\boldsymbol{z}$, the optimal solution of model (21) must be reached at an extreme point of the box uncertainty set $\mathbb{U}$ [35], and $\mathbb{U}$ has a finite number of extreme points.

According to Proposition 2 given by [37], the proposed PC&CG algorithm with the above properties is guaranteed to converge to the optimal value of the equivalent model in (19) within a finite number of iterations, and with the equivalent conditions given by (19), the PC&CG algorithm is also guaranteed to converge to the optimal value of model (18).

In addition, the proposed PC&CG algorithm mitigates the computational burden of the LCBDAR assessment model through parallel solving, the subproblem size of which is reduced by $N_L$ times compared to the traditional C&CG, thus reducing the computational time of the subproblems in each iteration, and its computational performance under different system scales is verified in Section VI.C.

## V. DECENTRALIZED SOLUTION BASED ON THE LCBDAR

### A. Noniterative Decentralized Solution Process

Based on the LCBDAR mentioned in Section IV, the following noniterative decentralized solution can be developed [2],[21]. For each RTED time window $T$:

1) The NGS operator solves the LCBDAR assessment model by the proposed PC&CG to obtain the LCBDAR $\mathcal{Q}$.

2) Through the electricity-gas conversion function (13), the LCBDAR $\mathcal{Q}$ that characterizes the secure gas withdrawal ranges for GFUs is converted into the feasible ranges of the real-time outputs $\tilde{\boldsymbol{P}}^{GF}$ for GFUs:

$$\mathcal{P} := \left\{ \tilde{\boldsymbol{P}}^{GF} \mid \underline{\boldsymbol{P}}^{GF} \leq \tilde{\boldsymbol{P}}^{GF} \leq \bar{\boldsymbol{P}}^{GF} \right\} \quad (22)$$

where $\bar{\boldsymbol{P}}^{GF}$ and $\underline{\boldsymbol{P}}^{GF}$ denote the vectors of the upper and lower bounds of the real-time outputs for the GFUs, which are converted from $\boldsymbol{q}^u$ and $\boldsymbol{q}^d$ through (13), respectively.

3) The converted feasible region $\mathcal{P}$ in (22) is then sent to the EPS. The EPS operator incorporates $\mathcal{P}$ into the RTED model (5a)-(5b), and establishes the following LCBDAR-constrained RTED (*LRTED*) model in the EPS:

$$\text{Obj: } \min_{\boldsymbol{x}_t^b, \boldsymbol{x}_t^s} \left\{ \sum_{t=1}^{N_\tau} \boldsymbol{c}_b^T \boldsymbol{x}_t^b + \frac{1}{N} \sum_{n=1}^{N} \sup_{\xi_t \in \mathcal{U}_N^\ell} \sum_{t=1}^{N_\tau} \boldsymbol{c}_s^T \boldsymbol{x}_t^s \xi_t \right\}$$

$$\text{s.t.} \begin{cases} \max_{\xi_t \in \mathcal{U}_N^\ell} \left( \boldsymbol{A}_t \boldsymbol{x}_t^s \xi_t + \boldsymbol{B}_t \boldsymbol{x}_{t-1}^s \xi_t \right) + \boldsymbol{C}_t \boldsymbol{x}_t^b + \boldsymbol{D} \boldsymbol{x}_{t-1}^b \leq \boldsymbol{a}_t, \ \forall t \in N_\tau \\ \boldsymbol{E} \boldsymbol{x}_t^s + \boldsymbol{F}_t \boldsymbol{x}_t^b = \boldsymbol{b}_t, \ \forall t \in N_\tau \\ \underline{\boldsymbol{P}}_t^{GF} \leq \tilde{\boldsymbol{P}}_t^{GF} \leq \bar{\boldsymbol{P}}_t^{GF}, \ \tilde{\boldsymbol{P}}_t^{GF} = \boldsymbol{K} \boldsymbol{x}_t^b + \boldsymbol{L} \boldsymbol{x}_t^s \xi_t, \ \forall t \in N_\tau \end{cases} \quad (23)$$

4) Solve the LRTED model (23) to obtain the RTED decisions $\boldsymbol{x}_t^b, \boldsymbol{x}_t^s$, which include the base outputs, reserves, and AGC participation factor of units. According to the discussion of LCBDAR in Section IV, within the adjustable range of the gas wells and compressors in the NGS, RTED decisions obtained by the LRTED model (23) can ensure the secure operation of the NGS under arbitrary pipeline leakage conditions, meanwhile avoiding the gas load shedding.

***Remark 4***: In the proposed decentralized solution, the "*noniterative*" means that there is no iteration between the EPS-side and the NGS-side model. Each RTED only requires a one-time boundary information exchange from the EPS to the NGS, thus protecting the system's inner data and avoiding the common problem of excessive iterations between EPSs and NGSs in many conventional decentralized algorithms.

### B. Robust Reformulation Method for Solving LRTED

Based on the robust sample path-based uncertainty model in Section III.A, the LRTED model (23) can be reformulated as the following robust optimization problem:

$$\min_{\boldsymbol{x}_t^b, \boldsymbol{x}_t^s} \left\{ \sum_{t=1}^{N_\tau} \boldsymbol{c}_b^T \boldsymbol{x}_t^b + \frac{1}{N} \sum_{n=1}^{N} \sup_{\xi_t \in \mathcal{U}_N^\ell} \sum_{t=1}^{N_\tau} \boldsymbol{c}_s^T \boldsymbol{x}_t^s \xi_t \right\}$$

$$\text{s.t. } \boldsymbol{A}_t^R \boldsymbol{x}_t^b + \boldsymbol{D}_t^R \boldsymbol{x}_t^s = \boldsymbol{a}_t^R, \ \boldsymbol{C}_t^R \boldsymbol{x}_t^b \leq \boldsymbol{c}_t^R, \ \forall t \in N_\tau \quad (24)$$

$$\sum_{t=1}^{N_\tau} \left( \boldsymbol{E}_t^R + \sum_{i=1}^{N_\tau} \boldsymbol{A}_i^R \boldsymbol{x}_i^s \right) \xi_t \leq \boldsymbol{b}_t^R + \sum_{t=1}^{N_\tau} \boldsymbol{B}_t^R \boldsymbol{x}_t^b, \ \forall \xi_t \in \mathcal{U}$$

where the constraints in (24) are derived from constraints (5b) and (22) through matrix transformation. The variables in (24) were defined in the above sections.

By adopting the strong duality [24], the robust optimization model (24) can be further reformulated as the following linear optimization model:

$$\min_{x_t^b, x_t^s} \left\{ c_b^T x_t^s + \frac{1}{N} \sum_{n=1}^{N} \sup_{\xi_t \in \mathcal{U}_N^\ell} \sum_{t=1}^{N_\tau} \left( \omega_t \cdot \vartheta_t^\ell - \upsilon_t \cdot \partial_t^\ell \right) \right\}$$

$$\text{s.t. } A_t^R x_t^b + D_t^R x_t^s = a_t^R, \; C_t^R x_t^b \le c_t^R, \; \forall t \in N_\tau$$

$$\sum_{t=1}^{N_\tau} \left[ \mu_t \cdot \vartheta_t^\ell - \eta_t \cdot \partial_t^\ell \right] \le b_t^R + \sum_{t=1}^{N_\tau} B_t^R x_t^b, \; \forall \ell \in \mathcal{N} \quad (25)$$

$$\mu_t - \eta_t = E_t^R + \sum_{i=1}^{N_\tau} A_i^R x_i^s, \; \omega_t - \upsilon_t = c_s^T x_t^s, \; \forall t \in N_\tau$$

where $\mu_t$, $\eta_t$, $\omega_t$ and $\upsilon_t$ denote the dual auxiliary variables.

According to Theorem 1 and Proposition 1 in [24], the solution of the linear optimization model (25) is guaranteed to converge to the optimal value of the original model (23). Due to the space limitation, the detailed proof is omitted here.

## VI. CASE STUDY

### A. Setup

Three test systems are used to verify the effectiveness of the proposed models and algorithms: a 22-bus EPS with a 10-node NGS (E22-N10), a 39-bus EPS with a 20-node NGS (E39-N20), and a modified 118-bus EPS with a 40-node NGS (E118-N40). The detailed parameters of these systems can be found in Ref. [9] and the electronic companion [31].

The following RTEDs are executed for $T=1$ hour with the temporal resolution set to $\tau = 5$ minutes. The historical wind data are sampled from a wind farm in northern China to obtain 100 wind power scenarios as in-sample data support for RTED optimization. By randomly weighting different in-sample data and adding Gaussian white noise, 500 wind generation scenarios are generated as the test set $\Psi$, which is used to test the out-of-sample performance of the proposed models and algorithms. The software *Pipeline Studio* is used for the dynamic state simulation of the pipeline network. In addition, to compare the performances of different models and algorithms, we introduce the following two metrics:

- *The pressure violation ratio* (PVR): the ratio of the number of pressure violation scenarios to the total number of wind scenarios in $\Psi$ under all NGS conditions in $\mathcal{W}$.
- *The reserve shortage ratio* (RSR): the ratio of the number of reserve shortage scenarios to the total number of wind scenarios in $\Psi$ under all NGS conditions in $\mathcal{W}$.

All the optimization models in the case studies are solved by Gurobi 10.0.1 and IPOPT in MATLAB R2022b on a computer with a 2.10 GHz CPU and 16 GB RAM.

### B. Comparison of different RTED models

To verify the effectiveness of the proposed RTED model, the following three models are employed on the three IEGSs and evaluated on the test set $\Psi$ under all the pipeline network conditions in $\mathcal{W}$:

- *Traditional RTED*: the traditional RTED model for the EPS, which optimizes RTED decisions without considering the NGS security.
- *Centralized RTED*: the proposed RTED model (14) in Section III.D, which optimizes RTED decisions considering the IEGS security constraints.
- *LRTED*: the LRTED model (23) in Section V.B, which optimizes RTED decisions considering the LCBDAR.

TABLE I
Dispatch and test results of different RTED models

| System | Solution | Traditional RTED | Centralized RTED | LRTED |
|---|---|---|---|---|
| E22-N10 | Cost ($10^7$\$) | 1.34 | 1.37 | 1.38 |
|  | RSR (%) | 0 | 0 | 0 |
|  | PVR (%) | 34 | 0 | 0 |
| E39-N20 | Cost ($10^7$\$) | 1.74 | 1.76 | 1.78 |
|  | RSR (%) | 0 | 0 | 0 |
|  | PVR (%) | 43.2 | 0 | 0 |
| E118-N40 | Cost ($10^7$\$) | 2.37 | No solution is obtained before 900 seconds | 2.38 |
|  | RSR (%) | 0 |  | 0 |
|  | PVR (%) | 40.1 |  | 0 |

According to the comparison results in Table I, the following three conclusions can be drawn:

First, The RSRs related to the three models in different IEGSs are all 0%, which verifies that these models all efficiently ensure the EPS-side operational security.

Second, the PVRs related to the traditional RTED in different IEGSs are high. In contrast, the PVRs associated with the centralized RTED are all 0%, meaning that the proposed RTED with the IEGS security constraints can ensure the NGS-side operational security, while decisions obtained by the traditional RTED without considering the NGS security will pose risks to NGS operations.

Third, similar to the centralized RTED, the PVRs related to the LRTED model in different IEGSs are also 0%, which verifies the effectiveness of the proposed LCBDAR. This is because the proposed NGS-side LCBDAR assessment model effectively considers all the worst-case gas withdrawal scenarios for GFUs under all potential pipeline leakage conditions when determining the LCBDAR that will be reliable in arbitrary wind power fluctuation and pipeline leakage scenarios in $\Psi \times \mathcal{W}$. Thus, LRTED can fully utilize the flexibility of NGS while ensuring the operational security of the IEGS and avoiding load shedding in the EPS and NGS.

Finally, the optimal costs of the LRTED model are only approximately 1% higher than those of the centralized RTED, this acceptable cost increase indicates that the LRTED has satisfactory economic performance. Besides, the optimal costs of the LRTED model are all higher than those of traditional RTED. This is because the proposed LRTED model ensures the reliability of the generation decisions of GFUs for both the EPS and the NGS, and optimizes decisions in a more secure manner than the RTED model.

To intuitively illustrate the impact of the LCBDAR on the security of RTED decisions, a simulation of a real-time dispatch scenario on the E22-N10 system is carried out, in which leakage failure occurs at the midpoint of the 9[th] pipeline, the wind generation scenario is randomly selected from the test set $\Psi$, the lower bound of the pressure in the NGS is set to 4 MPa, and the prices of GFU-1 to GFU-4 increase sequentially. The dynamic pressure fluctuations of the pipeline network are simulated by the software Pipeline Studio.

As indicated by the blue curves in Fig. 5, due to the higher price of GFU-4, the RTED without the LCBDAR allocates more regulating capacity to GFU-1 to GFU-3 based on economic optimality and decision reliability. This makes the output and corresponding gas withdrawal fluctuations of GFU-

2 and GFU-3 more significant, and the excessive gas withdrawal finally leads to pressure violations in the pipeline network, as shown by the blue curves in Fig. 6. After considering the LCBDAR, as indicated by the red curves in Fig. 5, the excessive gas withdrawals of GFUs that may cause pressure violations are reduced by LRTED; this prevents significant pressure fluctuations in the NGS, as shown by the red curves in Fig. 6, thus ensuring NGS operational security.

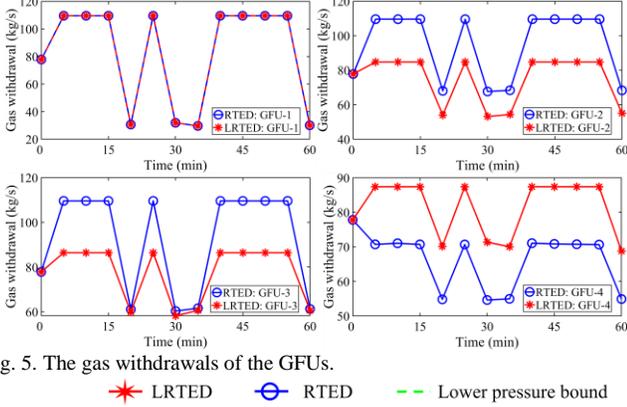

Fig. 5. The gas withdrawals of the GFUs.

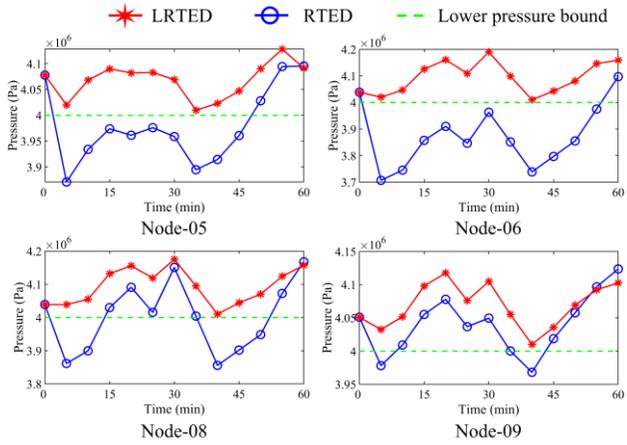

Fig. 6. The pressure fluctuations of the pipeline network.

## C. Computational Performance

### 1) Performance of the Customized PC&CG Algorithm and the Robust Reformulation Method

To test the computational performance of the customized PC&CG for the LCBDAR assessment model in Section IV.B and the robust reformulation for LRTED in Section V.B, under each IEGS, 10 different power demand and traditional gas load conditions are selected, and the PC&CG algorithm and the robust reformulation method is implemented 10 times under each condition to obtain the maximum, minimum, and average computational times, and numbers of iterations.

TABLE II
Performance of the customized PC&CG

| System | Computational time (second) | | |
|---|---|---|---|
| | Average | Minimum | Maximum |
| E22-N10 | 10.19 | 9.12 | 13.79 |
| E39-N20 | 48.20 | 39.49 | 58.11 |
| E118-N40 | 92.57 | 87.16 | 107.13 |
| System | Iterations[1] | | |
| | Average | Minimum | Maximum |
| E22-N10 | 3 | 2 | 5 |
| E39-N20 | 3 | 2 | 5 |
| E118-N40 | 4 | 2 | 5 |

[1] Number of iterations between the master problem and subproblems.

Table II presents the computational performance of the PC&CG algorithm across various IEGS configurations; the average computational times for all scenarios are less than 2 minutes, and the average number of iterations is less than 5. Given that the RTED is executed every 15 minutes, these test results of PC&CG underscore the satisfactory convergence and computational time of the proposed PC&CG algorithm. Moreover, the results highlight the good adaptability of this approach to different scales of IEGSs.

TABLE III
Performance of the robust reformulation method

| System | Computational time (second) | | |
|---|---|---|---|
| | Average | Minimum | Maximum |
| E22-N10 | 2.12 | 1.59 | 4.11 |
| E39-N20 | 6.80 | 5.70 | 7.42 |
| E118-N40 | 17.13 | 15.42 | 19.14 |

Table III illustrates that the computational times of the robust reformulation method in different IEGSs are within 20 seconds, indicating that this method has a satisfactory computational speed in solving the EPS-side RTED problems. In particular, the maximum and minimum computational times of the robust reformulation method differ by less than 2 seconds, indicating that this method has stable computational performance in different scales of IEGSs.

### 3) Performance of the Decentralized Solution

The overall computational performance of the proposed decentralized solution based on the LCBDAR is compared with that of the centralized solution obtained by the IPOPT solver and the conventional ADMM [38]. The comparisons are shown in Table IV.

TABLE IV
Optimal costs and solution times of different methods

| System | Method | Cost ($10^7$\$) | Solution time[1] (second) | Iterations between EPS and NGS |
|---|---|---|---|---|
| E39-N20 | Proposed solution | 1.78 | 54.97 | 1 |
| | Centralized solution | 1.76 | 774.09 | Not applicable |
| | ADMM | 1.77 | 472.89 | 7 |
| E118-N40 | Proposed solution | 2.38 | 108.73 | 1 |
| | Centralized solution | No solution is obtained before 900 seconds (15 minutes) | | |
| | ADMM | No solution is obtained before 900 seconds (15 minutes) | | |

[1] Sum of the computational times of PC&CG in NGS and LRTED in EPS.

In the E39-N20 system, the overall cost of the proposed decentralized solution is 1.13% greater than that of the centralized solution obtained by IPOPT, while the computational time of the proposed decentralized solution based on LCBDAR is only approximately 7% of that of the centralized solution obtained by IPOPT. This comparison highlights the advantage of the proposed decentralized solution in ensuring the privacy of the internal data of the EPS and NGS and the superior computational speed compared to that of the centralized solution obtained by IPOPT. Moreover, the computational time of the proposed decentralized solution is only approximately 11% of that of the conventional ADMM. This remarkable efficiency stems from the noniterative nature of the proposed decentralized solution, which eliminates the necessity of recurrently addressing subproblems related to the EPS and NGS.

In the larger E118-N40 system, neither the IPOPT nor the ADMM can solve the model within 15 minutes, a level of inefficiency that is deemed unacceptable for practical field applications. Conversely, the computational time required by the proposed decentralized solution is less than 2 minutes, thus reaffirming the superior performance of the proposed decentralized solution for a comparatively extensive IEGS.

## VII. Conclusion

This work presented an RTED model for IEGSs considering uncertainty propagation and pipeline leakage uncertainty and developed a detailed and efficient algorithm for assessing the LCBDAR for NGSs. Through the LCBDAR, the complicated NGS dynamic operational state constraints can be transformed into a set of intuitive, secure gas withdrawal and generation output ranges for GFUs, which can be easily incorporated into the RTED model for IEGSs. Moreover, the proposed noniterative decentralized solution based on the LCBDAR reduces the computational burden associated with the RTED of the IEGS while maintaining the high accuracy of decisions and preserving the information privacy of the EPS and NGS. The case studies demonstrate the superior efficacy of the proposed model and algorithms in the real-time optimal dispatch of IEGSs at different scales.

The proposed RTED with the LCBDAR facilitates the real-time collaboration of an IEGS with a high RES proportion and large-scale GFUs and can make full use of the flexibility provided by NGS to GFUs while ensuring the secure operation of the IEGS and avoiding load shedding in the EPS and NGS.

Future works will focus on 1) incorporating frequency constraints into the EPS-side RTED model to further ensure frequency security and 2) developing an LCBDAR for an IEGS with GFUs and power-to-gas to further exploit the flexibility of an NGS in the context of two-way coupling.

## References

[1] Y. Zhang, Z. Huang, F. Zheng, et al. "Cooperative optimization scheduling of the electricity-gas coupled system considering wind power uncertainty via a decomposition-coordination framework," *Energy*, vol. 194, Mar. 2020.

[2] S. Wang, W. Wu, C. Lin, et al. "A dynamic equivalent energy storage model of natural gas networks for joint optimal dispatch of electricity-gas systems," *IEEE Trans. Sustain. Energy*, vol.15, no. 1, pp. 621-632, Jan. 2024.

[3] Z. Liu, H. Li, K. Hou, et al. "Risk assessment and alleviation of regional integrated energy system considering cross-system failures," *Appl. Energy*, vol. 350, Nov. 2023.

[4] J. Yang, N. Zhang, C. Kang, et al. "Effect of natural gas flow dynamics in robust generation scheduling under wind uncertainty," *IEEE Trans. Power Syst.*, vol. 33, no. 2, pp. 2087-2097, Mar. 2018.

[5] C. Wang, R. Zhang, and T. Bi, "Energy management of distribution-level integrated electric-gas systems with fast frequency reserve," *IEEE Trans. Power Syst.*, vol. 39, no. 2, pp. 4208-4223, Mar. 2024.

[6] Y. Wang, K. Zhang, and K. Qu, "Segmented real-time dispatch model and stochastic robust optimization for power-gas integrated system with wind power uncertainty," *J. Mod. Power Syst. Clean Energy*, vol. 11, no. 5, pp. 1480-1493, Sept. 2023.

[7] B. Zhao, T. Qian, W. Li, et al. "Fast distributed co-optimization of electricity and natural gas systems hedging against wind fluctuation and uncertainty," *Energy*, vol. 298, Jul. 2024.

[8] F. Qi, M. Shahidehpour, F. Wen, et al. "Decentralized privacy-preserving operation of multi-area integrated electricity and natural gas systems with renewable energy resources," *IEEE Trans. Sustain. Energy*, vol. 11, no. 3, pp. 1785-1796, Jul. 2020.

[9] Z. Zhang, C. Wang, S. Chen, et al. "Multitime scale co-optimized dispatch for integrated electricity and natural gas system considering bidirectional interactions and renewable uncertainties," *IEEE Trans. Ind. Appl.*, vol. 58, no. 4, pp. 5317-5327, July-Aug. 2022.

[10] L. Yang, Y. Xu, J. Zhou, et al. "Distributionally robust frequency constrained scheduling for an integrated electricity-gas system," *IEEE Trans. Smart Grid*, vol. 13, no. 4, pp. 2730-2743, Jul. 2022.

[11] M. Bao, Y. Ding, C. Shao, et al. "Nodal reliability evaluation of interdependent gas and power systems considering cascading effects," *IEEE Trans. Smart Grid*, vol. 11, no. 5, pp. 4090-4104, Sept. 2020.

[12] G. Sun, S. Chen, Z. Wei, et al. "Corrective security-constrained optimal power and gas flow with binding contingency identification," *IEEE Trans. Sustain. Energy*, vol. 11, no. 2, pp. 1033-1042, Apr. 2020.

[13] C. He, L. Wu, T. Liu, et al. "Robust co-optimization planning of interdependent electricity and natural gas systems with a joint N-1 and probabilistic reliability criterion," *IEEE Trans. Power Syst.*, vol. 33, no. 2, pp. 2140-2154, Mar. 2018.

[14] W. Liang, S. Lin, M. Liu, et al. "Risk assessment for cascading failures in regional integrated energy system considering the pipeline dynamics," *Energy*, vol. 270, May. 2023.

[15] S. Zhang, et al. "Dynamic modeling and simulation of integrated electricity and gas systems," *IEEE Trans. Smart Grid*, vol. 14, no. 2, pp. 1011-1026, Mar.2023.

[16] W. Liang, S. Lin, M. Liu, et al. "Risk-based distributionally robust optimal dispatch for multiple cascading failures in regional integrated energy system using surrogate modeling," *Appl. Energy*, vol. 353, Jan. 2024.

[17] F. Shen et al. "Two-stage optimal dispatch of electricity-natural gas networks considering natural gas pipeline leakage and linepack," accepted by *IEEE Trans. Smart Grid*, doi: 10.1109/TSG.2024.3366943.

[18] S. Zhang et al. "Towards fast and robust simulation in integrated electricity and gas system: a sequential united method," *IEEE Trans. Power Syst.*, vol. 39, no. 1, pp. 1822-1836, Jan. 2024.

[19] W. Lin, X. Jin, H. Jia, et al. "Decentralized optimal scheduling for integrated community energy system via consensus-based alternating direction method of multipliers," *Appl. Energy*, vol. 302, Nov. 2021.

[20] M. Yan, W. Gan, Y. Zhou, et al. "Projection method for blockchain-enabled non-iterative decentralized management in integrated natural gas-electric systems and its application in digital twin modeling," *Appl. Energy*, vol. 311, Apr. 2022.

[21] H. Peng, M. Yan, and Y. Zhou, "Privacy-preserving non-iterative decentralized optimal energy flow for integrated hydrogen-electricity-heat system based on projection method," *Appl. Energy*, vol. 368, Aug. 2024.

[22] M. Farrokhifar, Y. Nie, and D. Pozo, "Energy systems planning: A survey on models for integrated power and natural gas networks coordination," *Appl. Energy*, vol. 262, Mar. 2020.

[23] Z. Li, W. Wu, B. Zhang, et al. "Adjustable robust real-time power dispatch with large-scale wind power integration," *IEEE Trans. Sustain. Energy*, vol. 6, no. 2, pp. 357-368, Apr. 2015.

[24] D. Bertsimas, S. Shtern, and B. Sturt. "A data-driven approach to multistage stochastic linear optimization," *Manage. Sci.*, vol. 69, no. 1, pp. 51-74, Mar. 2022.

[25] K. Qu, X. Zheng, X. Li, et al. "Stochastic robust real-time power dispatch with wind uncertainty using difference-of-convexity optimization," *IEEE Trans. Power Syst.*, vol. 37, no. 6, pp. 4497-4511, Nov. 2022.

[26] H. Qiu, et al., "A historical-correlation-driven robust optimization approach for microgrid dispatch," *IEEE Trans Smart Grid*, vol. 12, no. 2, pp. 1135-1148, Mar. 2021

[27] Y. Zhang, Z. Huang, F. Zheng, et al. "Cooperative optimization scheduling of the electricity-gas coupled system considering wind power uncertainty via a decomposition-coordination framework", *Energy*, vol.194, Mar.2020.

[28] Y. Zhou, Z. Li, and G. Wang, "Study on leveraging wind farms' robust reactive power range for uncertain power system reactive power optimization," *Appl. Energy*, vol. 298, Sept. 2021.

[29] K. Qu, Y. Chen, S. Xie, et al. "Segmented distributionally robust optimization for real-time power dispatch with wind uncertainty," *IEEE Trans. Power Syst.*, vol. 39, no. 2, pp. 2970-2983, Mar. 2024.

[30] Y. Su, F. Liu, Z. Wang, et al. "Multi-stage robust dispatch considering demand response under decision-dependent uncertainty," *IEEE Trans. Smart Grid*, vol. 14, no. 4, pp. 2786-2797, Jul. 2023.

[31] "Stochastic real-time economic dispatch for integrated electric and gas systems considering uncertainty propagation and pipeline leakage" [Online]. Available: https://github.com/PeiyaoZhao01/Electronic-Companion.

[32] Y. Zhang, Y. Liu, S. Shu, et al. "A data-driven distributionally robust optimization model for multi-energy coupled system considering the temporal-spatial correlation and distribution uncertainty of renewable energy sources," *Energy*, vol. 216, Feb. 2021.

[33] J. H. Ferziger, M. Perić, and R. L. Street. Computational methods for fluid dynamics[M]. *Springer*, 2019.

[34] Q. Chen, X. Xing, C. Jin, et al. "A novel method for transient leakage flow rate calculation of gas transmission pipelines," *J. Nat. Gas Sci. Eng.*, vol. 77, May. 2020.

[35] X. Chen, and N. Li. "Leveraging two-stage adaptive robust optimization for power flexibility aggregation," *IEEE Trans. Smart Grid*, vol. 12, no. 5, pp. 3954-3965, Sept. 2021.

[36] Z. Li, J. Wang, H. Sun, et al. "Robust estimation of reactive power for an active distribution system," *IEEE Trans Power Syst*, vol. 34, no. 5, pp. 3395-3407, Sept. 2019.

[37] B. Zeng, and L. Zhao. "Solving two-stage robust optimization problems using a column-and-constraint generation method," *Oper. Res. Lett.*, vol. 41, no. 5, pp. 457-461, Sept. 2013.

[38] S. Boyd, N. Parikh, E. Chu, et al. "Distributed optimization and statistical learning via the alternating direction method of multipliers," *Found. Trends Mach. Learn.*, vol. 3, no. 1, pp. 1-122, Jul. 2011.